\documentclass[preprint,review,12pt]{elsarticle}

\usepackage{amssymb}

\usepackage{lineno}

\journal{Ultramicroscopy}

\begin{document}

\begin{frontmatter}



\title{Magnetic properties of single nanomagnets: EMCD on FePt nanoparticles}


\author[IFW,IFP]{Sebastian Schneider\fnref{IFWadress}}
\ead{sebastian.schneider@ifw-dresden.de}

\author[IFW]{Darius Pohl}

\author[USTEM,MCMASTER] {Stefan L\"{o}ffler}

\author[UUPPSALA] {J\'{a}n Rusz}

\author[MPICPfS] {Deepa Kasinathan}

\author[USTEM,TUWIEN] {Peter Schattschneider}

\author[IFW,IFP]{Ludwig Schultz}

\author[IFW]{Bernd Rellinghaus}

\fntext[IFWadress]{Corresponding author. Tel.: +49 351 4659 640;
Fax: +49 351 4659 9754.}

\address[IFW]{IFW Dresden, Institute for Metallic Materials, PO Box 270116, D-01171 Dresden, Germany}
\address[IFP]{TU Dresden, Institut f\"{u}r Festk\"{o}rperphysik, D-01062 Dresden, Germany}
\address[USTEM]{TU Wien, University Service Centre for Electron Microscopy, Wiedner Hauptstra{\ss}e 8-10, A-1040 Vienna, Austria}
\address[MCMASTER]{McMaster University, Department of Materials Science and Engineering, 1280 Main Street West, Hamilton, Ontario L8S4L8, Canada}
\address[UUPPSALA]{Uppsala University, Department of Physics and Astronomy, PO Box 516, S-75120 Uppsala, Sweden}
\address[MPICPfS]{Max-Planck-Institut f\"{u}r Chemische Physik fester Stoffe, Department of Physics of Correlated Matter, N\"{o}thnitzer Stra{\ss}e 40, D-01187 Dresden, Germany}
\address[TUWIEN]{TU Wien, Institute of Solid State Physics, Wiedner Hauptstra{\ss}e 8-10, A-1040 Vienna, Austria}

\begin{abstract}
Energy-loss magnetic chiral dichroism (EMCD) allows for the quantification of magnetic properties of materials at the nanometer scale. It is shown that with the support of simulations that help to identify the optimal conditions for a successful experiment and upon implementing measurement routines that effectively reduce the noise floor, EMCD measurements can be pushed towards quantitative magnetic measurements even on individual nanoparticles. With this approach, the ratio of orbital to spin magnetic moments for the Fe atoms in a single L$1_0$ ordered FePt nanoparticle is determined to be ${m_l}/{m_s} = 0.08 \pm 0.02$. This finding is in good quantitative agreement with the results of XMCD ensemble measurements.
\end{abstract}


\end{frontmatter}


\section{Introduction}
\label{Introduction}

Transmission electron microscopy (TEM) has established itself as a, if not {\em the} standard method for studying materials at the atomic scale over the last decades. The variety of operational modes of a TEM allows for the quantitative investigation of the structure (high-resolution transmission electron microscopy (HRTEM) \citep{Urban2008}, scanning transmission electron microscopy (STEM) \citep{Muller2009}), the chemical composition or binding (electron energy loss spectroscopy (EELS) \citep{Egerton2009}, energy-dispersive X-ray spectroscopy (EDX) \citep{DAlfonso2010}) and even of the magnetic properties of a specimen (Lorentz microscopy \citep{Freitag2009}, holography \citep{Lichte2013}). However, the latter two modes still lack sub-nanometer resolution \citep{Dunin-Borkowski2012}. In times of ongoing miniaturization of data storage entities and attempts to implement spin degrees of freedom into electronics, the demand of a technique that would allow for the acquisition of magnetic information such as an the spin configuration quantitatively on the nanometer scale is beyond all questions. Meanwhile, energy-loss magnetic chiral dichroism (EMCD), which is the electron wave analogue of the already well established approach of X-ray magnetic circular dichroism (XMCD) \citep{Stohr1998}, has been proposed as a novel technique that promises to improve the spatial resolution of magnetic measurements in a TEM \citep{Hebert2003}. EMCD in combination with using electron vortex beams has the potential to push this envelope even further towards quantitative magnetic characterization with atomic resolution \citep{Verbeeck2010,Pohl2015}.

By now, `classical' EMCD, where the sample is conventionally illuminated with plane electron waves, was successfully demonstrated to reveal a dichroic signal for the magnetic characterization of extended ferromagnetic thin films with lateral resolutions in the nanometer regime \citep{Schattschneider2008a}. It even allows for the element-specific and/or site-specific extraction of magnetic moments \citep{Lidbaum2009,Loukya2012,Wang2013,Muto2014,Tatsumi2014,Thersleff2015,Fu2015,Song2015b}, thereby largely surpassing the resolution of XMCD experiments. Nevertheless, the acquisition of EMCD spectra from individual nanoparticles was so far scarcely reported (see, e.g., \citep{Salafranca2012}), and did not provide any quantitative information on the magnetic moments. This is mainly due to the difficulties with the sample stability and the (beam-induced) tilting of particles with dimensions of just a few $\mathrm{nm}$.

In the present paper, `classical' EMCD is used to investigate individual $\mathrm{L1_{0}}$ FePt nanoparticles \citep{Pohl2011,Pohl2014,Wicht2015}.  $\mathrm{L1_{0}}$ ordered FePt is a promising materials candidate for the media in future heat assisted magnetic recording, since it offers the highest magneto-crystalline anisotropy among the oxidation-resistant hard magnets \citep{Kryder2008,Wu2013,Lyubina2011}. We report on quantitative EMCD measurements on individual FePt nanoparticles and compare our experimental findings with simulations. The paper is divided into two main parts following the course of the conducted work. At first simulations on $\mathrm{L1_{0}}$ FePt are performed to find the optimum parameter space for EMCD experiments. In a second step the actual FePt nanoparticles are studied in a FEI Titan$\mathrm{^3}$ 80-300 microscope equipped with an image $\mathrm{C_S}$ corrector and a Gatan Tridiem 865 energy filter.

\section{Simulations}

\subsection{Optimal sample thickness for EMCD experiments on FePt}

In order to optimize the experimental conditions, simulations of the \linebreak EMCD effect are performed. Here, the elastic scattering of the probe electrons both before and after an inelastic scattering event are taken into account using Bloch waves calculated for the experimental scattering geometry, while the inelastic scattering is modeled using the mixed dynamic form factor (MDFF) approach \citep{Kohl1985,Schattschneider2000,Loffler2010}.

\begin{figure}[p]
        \centering
        \includegraphics{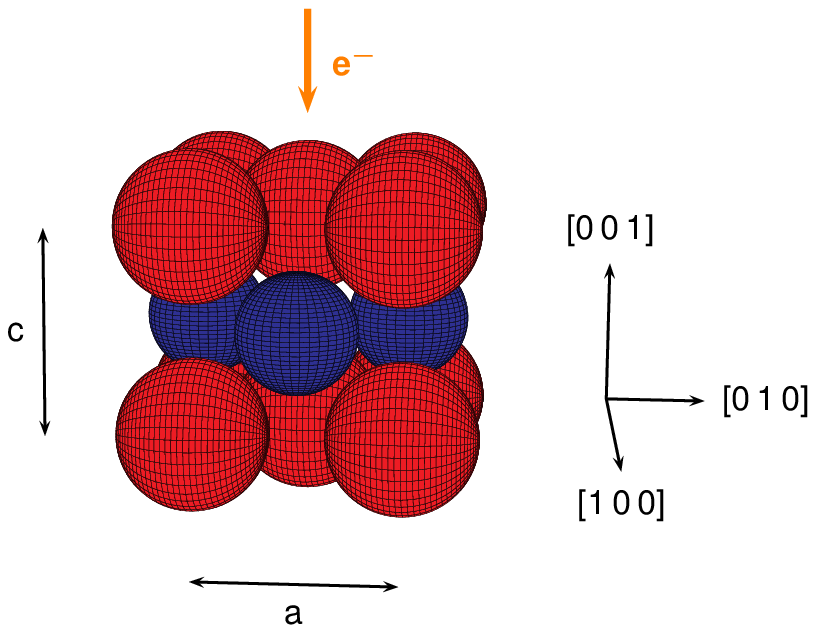}
        \caption{Unit cell of FePt with Fe atoms in red and Pt in blue. The beam does not traverse exactly along the $\left[ 0 \, 0 \, 1 \right]$ zone axis, but rather it is slightly tilted. The lattice parameters of FePt are $a = 3.86$ {\AA} and $c = 3.71$ {\AA}.}
        \label{FePt}
\end{figure}

Since the $\mathrm{L1_{0}}$ FePt crystals are grown with their [001] easy axis of magnetization perpendicular to the substrate, the electron beam lies parallel to this easy axis in the here investigated plan view samples. Consequently and as depicted in Fig. \ref{FePt}, the electrons are scattered at alternating layers of Fe and Pt atoms while traversing the crystal. In order to be able to pick up the EMCD, the sample is tilted by $3^\circ$ into the three beam case, with the center of the Laue circle shifted from $\left( 0 \, 0 \, 0 \right)$ to $\left( 0 \, 20 \, 0 \right)$ thereby strongly exciting the diffraction vectors $\vec{G} = \left( 2 \, 0 \, 0 \right)$ and $-\vec{G} = \left( \overline{2} \, 0 \, 0 \right)$, respectively. Then the positions `$++$'$= \left( 1 \, 1 \, 0 \right)$ and `$-+$'$= \left( \overline{1} \, 1 \, 0 \right)$, which lie on Thales circles through the diametrical pairs of positions $\vec{O}$, $\vec{G}$ and $\vec{O}$, $-\vec{G}$ in the diffraction plane, are chosen as the detector positions for the simulations. All calculations are performed with three incoming and three outgoing Bloch waves.

\begin{figure}[p]
        \centering
        \includegraphics[width=0.8\textwidth]{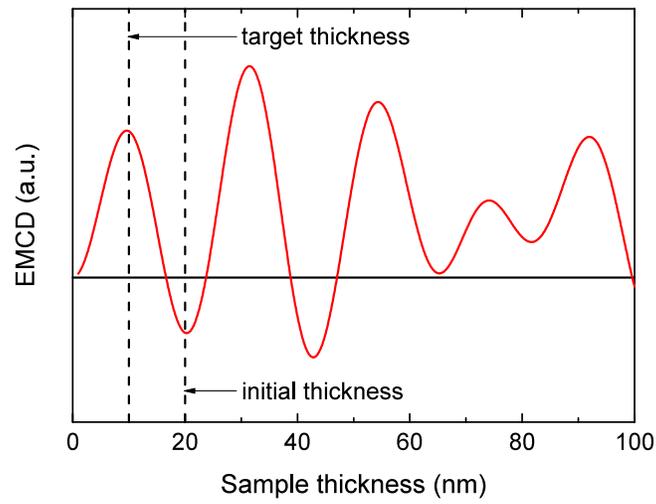}
        \caption{Dependence of the EMCD signal on the thickness of the FePt sample at the Fe $L_3$ edge $\left( E_{L_3} = 708 \ \mathrm{eV} \right)$.}
        \label{Thicknessmap}
\end{figure}

Fig. \ref{Thicknessmap} shows the oscillatory behaviour of the EMCD signal at the Fe $L_3$ absorption edge $\left(E_{L_3} = 708 \, \mathrm{eV} \right)$, stemming from the pendell\"{o}sung. A high dichroic signal is obtained at a sample thickness of $9.6 \, \mathrm{nm}$, whereas the signal is considerably smaller at $20.3 \, \mathrm{nm}$, which is roughly the initial thickness of the FePt nanoparticles. These simulations suggest that further thinning of the particles is necessary to be able to measure a sufficiently strong dichroic signal.


\subsection{EMCD map for aperture positioning}

Unlike in the above presented initial simulations, the detector entrance apertures are circular openings with finite diameters rather than being point-like. It is thus of utmost importance to know the momentum dependence of the EMCD signal, in order to be able to position the detector optimally in the diffraction plane during the experiment. Accordingly, EMCD maps are simulated in the momentum-dispersive diffraction plane. The code utilized for these calculations \citep{Rusz2013} is based on the same assumptions that were already used in the first simulation, however, with a larger basis set of now approximately 420 Bloch waves.

\begin{figure}[p]
        \centering
        \includegraphics[width=\textwidth]{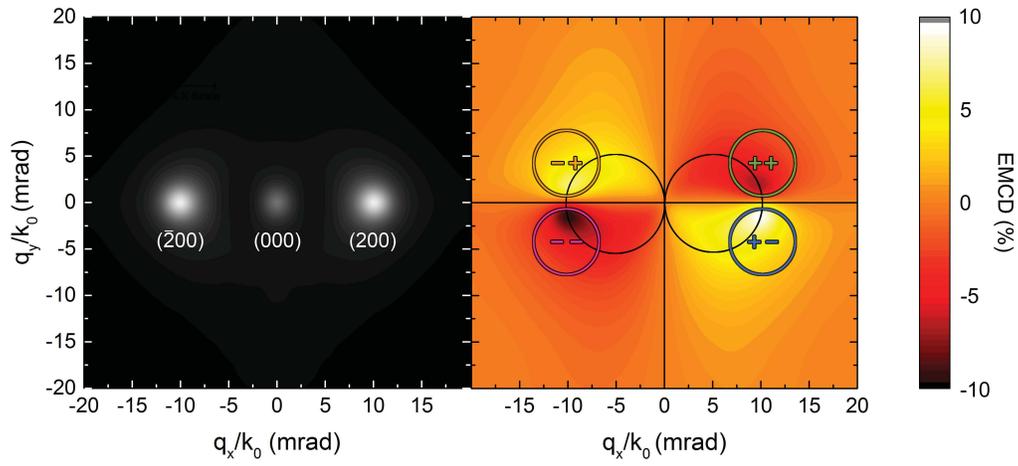}
        \caption{Simulated diffraction pattern (left) and EMCD map (right) at the Fe $L_3$ edge $\left( E_{L_3} = 708 \ \mathrm{eV} \right)$. The two Thales circles are depicted in black in the EMCD map. The four aperture positions used in the experiment are marked by coloured circles in the map.}
        \label{EMCD_map}
\end{figure}

Upon varying the sample thickness, the simulations also reveal a maximum of the EMCD signal for a thickness of $10 \, \mathrm{nm}$, which is in good agreement with the first calculation. The simulated diffraction pattern and EMCD map of a $\mathrm{L1_{0}}$ FePt crystal that is tilted out of the zone axis by $10^{\circ}$ are displayed in Figure \ref{EMCD_map}. Tilting of the sample results in a three beam case, with the first diffraction orders $\left( 2 \, 0 \, 0 \right)$ and $\left(\overline{2} \, 0 \, 0 \right)$ carrying a higher intensity than $\left( 0 \, 0 \, 0 \right)$. Such differences in the intensities of the diffraction spots are also a fingerprint of the correct thickness of the sample and can be compared to the experimental diffraction pattern.

The right panel of Fig. \ref{EMCD_map} shows the resulting dependence of the dichroic signal at the Fe $L_3$ absorption edge $\left(E_{L_3} = 708 \, \mathrm{eV} \right)$ on the position in the diffraction plane. The color scale of the map represents the strength of the EMCD signal. The two Thales circles are displayed in black. The colored circles represent the actual size of the spectrometer entrance aperture in the experiment. They are labeled with the signs of the $k_x$ and $k_y$ components of their origins in the diffraction plane, with the $\left( 0 \, 0 \, 0 \right)$ spot being the origin of the coordinate system. Similar to the observations of Thersleff et al.\ \citep{Thersleff2015}, the EMCD effect is found to be most pronounced in wider areas next to the $1^{st}$ order diffraction spots rather than at symmetric positions between $\vec{O}$ and $\pm\vec{G}$ (cf.\ the simulated diffraction pattern on the left panel in Fig. \ref{EMCD_map}). Positioning the spectrometer symmetrically between the diffraction spots would thus lead to a loss in signal strength. Hence, during the experimental EMCD measurements on the $\mathrm{L1_{0}}$ FePt nanoparticles, the detector entrance apertures are placed close to the $\left( 2 \, 0 \, 0 \right)$ and $\left(\overline{2} \, 0 \, 0 \right)$ diffraction spots on the Thales circles in order to pick up a maximal magnetic signal \citep{Verbeeck2008}.

\subsection{Simulation of EEL spectra}

In a third step, EEL spectra of L$1_0$ FePt are calculated in the energetic vicinity of the Fe $L_3$ and $L_2$ absorption edges. For this and based on the results of the preceding simulations, the calculations are performed for the optimal sample thickness at aperture positions `$++$' and `$-+$' in the diffraction plane. Then the difference of these spectra provides for the EMCD spectrum to be expected.


The spin-polarized material properties of FePt are calculated by means of {\it ab initio} density functional theory (DFT) methods. Simulations are performed using the full-potential augmented-plane-wave code {\tt WIEN2k}, \citep{Blaha2001} with the standard unit cell as well as with a supercell taking into account the core-hole effect. The plane-wave cutoff parameter $R_{\mathrm{MT}}K_{\mathrm{max}}$ is set to 7.0, while the atomic sphere radii ($R_{\mathrm{MT}}$) for Fe and Pt are set to 2.45 and 2.50 respectively. The local density approximation (LDA) with the Perdew and Wang flavor of the exchange and correlation potential \citep{Perdew1992} is chosen for the spin-polarized plus spin-orbit coupled calculations. A convergency of 10$^{-6} \, \mathrm{Ry}$ is achieved with 3000 $k$-points for the standard unit cell and with 500 $k$-points for the $2\times2\times2$ supercell. ELNES (electron loss near edge structure) spectra are then calculated using the program {\tt TELNES} \citep{Blaha2001,Nelhiebel1999,Jorissen2007}. Effects of a core-hole are modeled by removing one core electron and adding it to the valence, thereby preserving the charge neutrality within the unit cell.

\begin{figure}[p]
        \centering
        \includegraphics[width=0.8\textwidth]{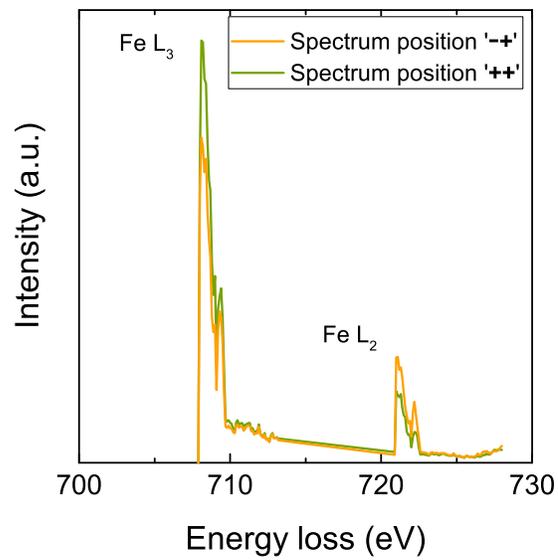}
        \caption{Simulated EEL spectra at the spectrometer positions `$-+$' and `$++$'. The two spectra show an asymmetry at the Fe absorption edges, which reverses from the $L_3$ to the $L_2$ edge. The lines in between the absorption edges ($713 - 720 \ \mathrm{eV}$) are just the connection lines between the last data point of the Fe $L_3$ and the first data point of the Fe $L_2$ edge.}
        \label{EEL_spectra_sim}
\end{figure}

Using the {\tt TELNES.3} routines included in {\tt WIEN2k}, the spin-polarized cross density of states (XDOS) and the radial wave functions of both the initial and final states of the target electron are extracted. Based on this data, the MDFF is calculated \citep{Schattschneider2006} for each energy loss of interest and used together with the elastic scattering calculations of the first simulation to compute the energy loss spectra at the detector positions used in the experiment. As can be seen from Figure \ref{EEL_spectra_sim}, the expected dichroism at the Fe absorption edges is indeed predicted from these calculations.


\section{Experimental methods}

\subsection{Preparation of the FePt nanoparticles}

FePt films with a thickness of roughly $15 \, \mathrm{nm}$ are prepared on single-crystalline SrTiO$_3$ (STO) substrates by co-sputtering from elemental targets in confocal geometry. Simultaneous heating of the rotating substrate holder leads to the formation of highly $\left[ 0 0 1 \right]$ textured L$1_0$ ordered FePt on STO. The elevated substrate temperatures induce de-wetting of the chemically ordered film and the subsequent formation of nanoparticulate FePt islands with thicknesses ranging from 14 to $20 \, \mathrm{nm}$ \citep{Thompson2012}.

\begin{figure}[p]
        \centering
        \includegraphics[width=0.6\textwidth]{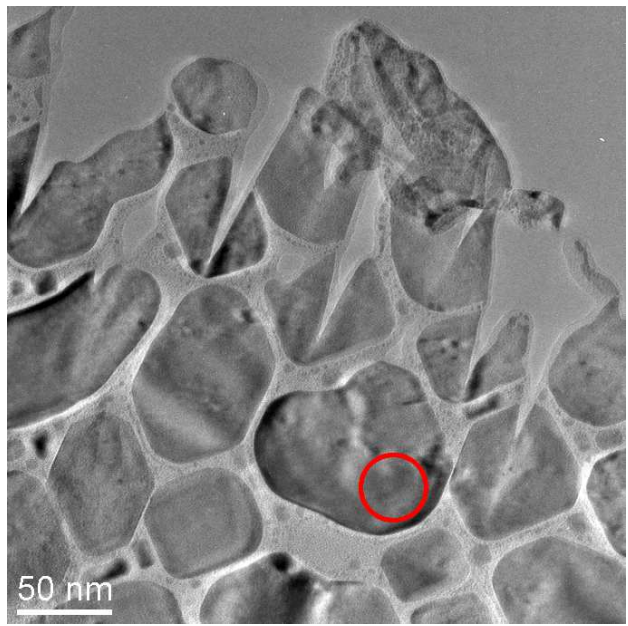}
        \caption{TEM image of thinned FePt nanoparticles. The red circle marks the area which is used for the spectroscopic measurements.}
        \label{FePt_nanoparticles}
\end{figure}

In order to provide for optimal conditions for the EMCD measurements, the FePt nanoparticles are further thinned to reach the desired target thickness of $10 \, \mathrm{nm}$. For this, the STO substrate is mechanically thinned to approximately $30 \, \mathrm{\mu m}$. In a second step, the sample is subjected to grazing incidence Ar$^{+}$ ion milling until a hole is formed. At the edge of the hole, free standing FePt nanoparticles can be found, which are solely held together by the glue introduced in the first preparation step. Figure \ref{FePt_nanoparticles} shows an overview plan view TEM image of the free standing L$1_0$ ordered FePt nanoparticles. No sign of Ti can be detected with EELS in this part of the sample thereby indicating that all of the STO has been removed.

\subsection{Adjusting the three beam case}

EMCD measurements are now conducted on one of the likewise prepared FePt nanoparticles. For this, the electron beam is condensed on the area marked by the red circle in Figure \ref{FePt_nanoparticles} resulting in a convergence angle of $0.4 \, \mathrm{mrad}$ (which provides for almost parallel illumination). Using the $\alpha$ and $\beta$ tilts of the sample holder, the orientation of the nanoparticle relative to electron beam is carefully changed, until the three beam case is adjusted as evidenced from the diffraction pattern. In order to provide for an orientation that is comparable to the one used in the simulations, the sample is adjusted until the intensity of the central $\left( 0 \, 0 \, 0 \right)$ spot is smaller than that of the $\left( 2 \, 0 \, 0 \right)$ and $\left(\overline{2} \, 0 \, 0 \right)$ $1^{st}$ order diffraction spots, respectively.

Since complete thickness series of EMCD maps and diffraction patterns are calculated, the diffraction pattern is used in addition to STEM-EELS-based thickness measurements to confirm the desired target sample thickness of $10 \, \mathrm{nm}$.

\begin{figure}[p]
        \centering
        \includegraphics[width=0.6\textwidth]{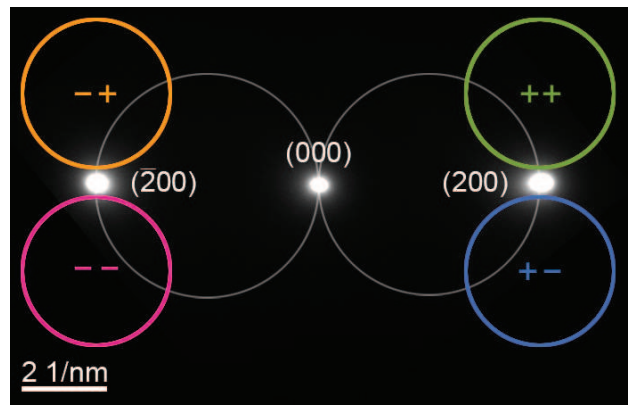}
        \caption{Experimental diffraction pattern of FePt in three beam case. The Thales circles are marked in grey and the four detector positions are also displayed as coloured circles.}
        \label{three_beam_case}
\end{figure}

The experiments are performed at an acceleration voltage of $300 \, \mathrm{kV}$. At the chosen camera length of $91 \, \mathrm{mm}$, the $1 \, \mathrm{mm}$ spectrometer entrance aperture provides for a collection semi-angle of $1.7 \, \mathrm{mrad}$. The aperture size and positions are indicated by colored circles in the diffraction pattern in Figure \ref{three_beam_case}. Only electrons from these areas of the diffraction pattern are guided into the spectrometer and contribute to the EEL spectra. The positioning is achieved by shifting the diffraction pattern relative to the spectromter entrance aperture with the help of the software tool {\tt loopy} \citep{Otten2013}, which allows to set, store and recall the desired positions within milliseconds. This effectively helps to limit the exposure time of the sample to the electron beam thereby preventing contamination \citep{Mitchell2015} and beam damage \citep{Egerton2004}.

\subsection{Measurement of the EEL spectra}

\begin{figure}[p]
        \centering
        \includegraphics[width=0.55\textwidth]{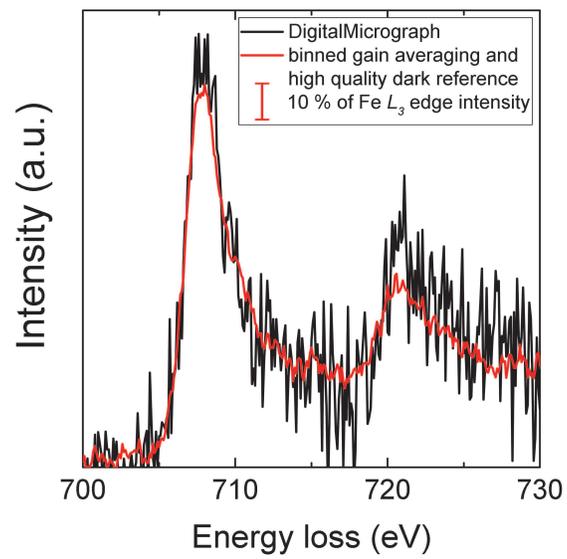}
        \caption{Comparison between sum spectra consisting of 100 single spectra with an acquisition time of $1.0 \, \mathrm{s}$ in EMCD geometry. The black line results from the standard {\tt DigitalMicrograph} measurement routine, whereas the red line is acquired with the improved method.}
        \label{binned_gain_averaging}
\end{figure}

Since the chosen detector positions lie in the diffuse background of the diffraction pattern and are well separated from areas with high electron intensities (namely the diffraction spots), the signal-to-noise ratio of the obtainable spectra is expected to be poor \citep{Verbeeck2008}. Hence, in order to not only be able to observe a qualitative EMCD effect, but also provide for the possibility to extract quantitative information, the noise level of the spectra has to be lowered significantly. For this purpose, three measures of improvement are implemented into the standard measuring process (see black line in Figure \ref{binned_gain_averaging}). (i) The binned gain averaging method is used to reduce the effect of correlated noise of the CCD camera \citep{Bosman2008}. (ii) Further improvement is achieved by subtracting a high quality (rather than standard) dark reference, by which the inherent correlated noise of the dark reference is reduced \citep{Hou2009}. (iii) Due to the insertion of the spectrometer entrance aperture, only a part of the CCD is exposed to electrons. Hence, reading out the whole CCD chip would add noise to the signal that originates from unexposed areas, which are inherently dark, but would add noise upon read-out. Accordingly, only the truely exposed part of the CCD is read out to collect the spectral information.

As can be seen from the red line in Figure \ref{binned_gain_averaging}, applying this advanced measurement routine allows for an acquisition of spectra with improved signal-to-noise ratio (SNR).

\begin{figure}[p]
        \centering
        \includegraphics[width=\textwidth]{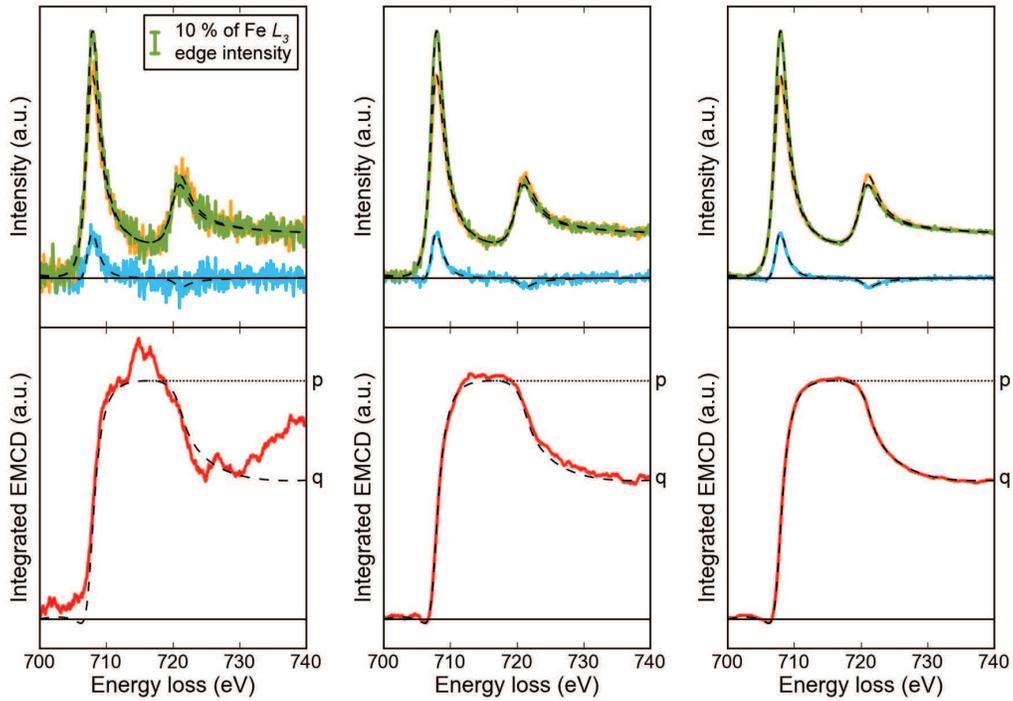}
        \caption{From left to right: Two EEL sum spectra (green and orange) consisting of 10, 100 and 1000 simulated single spectra with white and shot noise and their corresponding EMCD (blue) and integrated EMCD signal (red). The EEL spectra are normalised to the maximum intensity at the Fe $L_3$ absorption edge. For reference the simulated noisefree data are displayed as dashed black lines.}
        \label{noise}
\end{figure}

In order to (i) better understand the impact of noise and (ii) anticipate, at which SNR an extraction of quantitatively reliable EMCD spectra will become feasible, the impact of noise is firstly modeled. For this, an `ideal' spectrum is modeled by the sum of Gaussian, Lorentzian, step functions \citep{Lidbaum2010} and a power law background to mimic a spectrum as acquired at detector position `$-+$' with an acquisition time of $1.0 \, \mathrm{s}$. Gaussian noise with the correct amplitude to simulate the white noise of the CCD camera and Poisson noise are added to the spectrum to roughly match the experimental data shown in in Figure \ref{binned_gain_averaging}. At last, the background is removed by subtracting the power law function from the first step. The likewise simulated ELNES data is then comparable with the experimental data, from which the EMCD spectra are determined, due to the removal of the correlated noise using the advanced measurement procedure, only the random noise of the camera and the shot noise of the counting process contribute to the experimental spectra. Then, a second EEL spectrum at position `$++$' is simulated likewise. The averages over 10, 100 and 1000 of such pairs of noisy spectra are shown in the energetic vicinity of the Fe $L_3$ and $L_2$ absorption edges in Figure \ref{noise}. The noise-free spectra and the resulting EMCD signals are also displayed as dashed black lines in the top panel of Figure \ref{noise}. Integration of the EMCD signal over an energy window that covers the absorption edges results in the `perfect' integrated EMCD signal (dashed lines in the bottom left panel of Figure \ref{noise}). This integrated EMCD signal is of particular importance for quantifying magnetic properties from the EMCD. With the maximum value of the integrated EMCD after integrating the dichroism at the Fe $L_3$ edge, $p$, and the value $q$ of a plateau that forms after also integrating the dichroic signal of the Fe  $L_2$ edge (see bottom panel of Figure \ref{noise}) the ratio of the orbital to spin magnetic moments can be calculated to be $\frac{m_l}{m_s} = \frac{2q}{9p - 6q}$ according to the sum rules \citep{Rusz2007}.

After the addition of noise to the two perfect single EEL spectra for `$-+$' and `$++$', the original asymmetry between the two spectra is no longer recognizable (not shown here for the single spectra). However, upon averaging 10, 100, and 1000 of such noisy spectra (from left to right in Figure \ref{noise}) the SNRs in the sum spectra increase significantly (orange and green lines in the top panel of Figure \ref{noise}). And accordingly, the noise levels of the EMCD and integrated EMCD signals (blue and red lines in Figure \ref{noise}) decrease. While the average of 10 EEL spectra still reveals significant discrepancies to the noise-free results, the average over 100 spectra allows to extract an integrated EMCD signal that is already in good agreement with the data from the noise-free spectra. Averaging 1000 single EEL spectra results in an almost complete suppression of any noise artefacts, but for individual acquisition times of $1.0 \, \mathrm{s}$, such a high number of spectra is not suitable, as the total exposure times are limited due to the sample stability in the experiment.

\begin{figure}[p]
        \centering
        \includegraphics[width=0.75\textwidth]{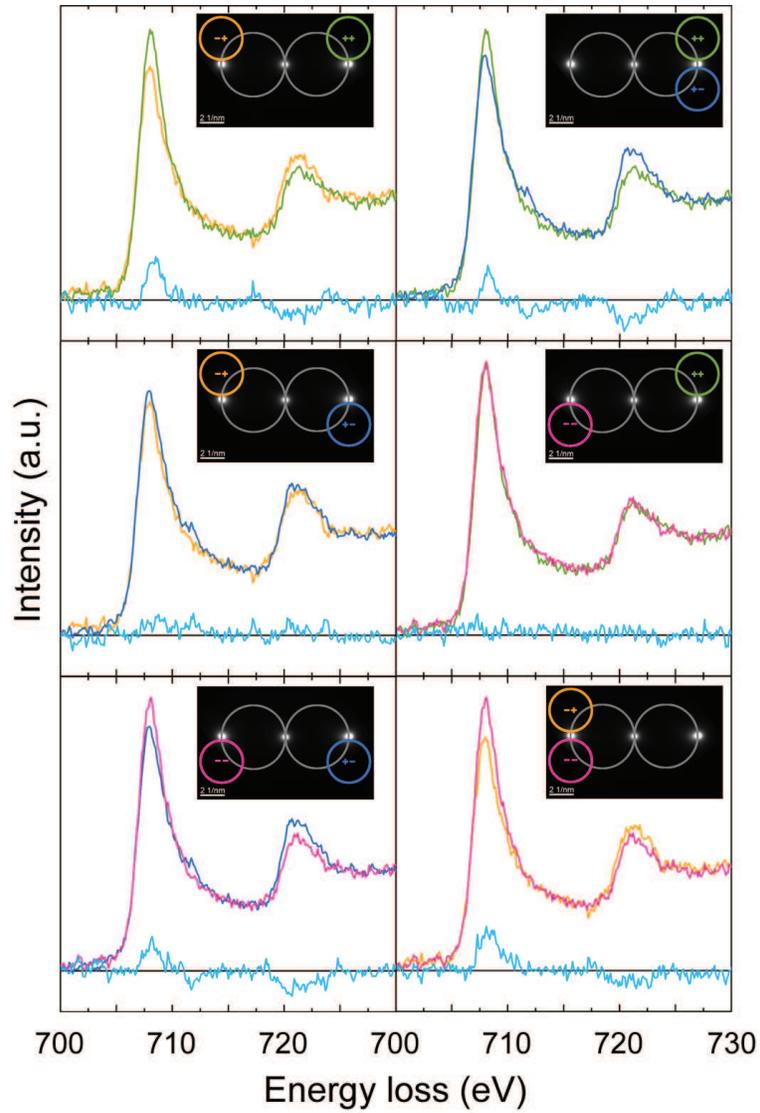}
        \caption{EEL spectra measured at the four detector positions with their respective EMCD signal (difference) in blue. The two aperture positions corresponding to the spectra are displayed as inset. Measurements with opposite polarisation of the electron wave (top and bottom row) show a dichroic signal, whereas the asymmetry at the absorption edges disappears when spectra are measured with the same polarisation of the electrons (middle row).}
        \label{EEL_spectra_exp}
\end{figure}

Hence based on these findings from the simulations of the noise effects, the experimental measurements are comprised of 100 spectra with individual acquisition times of $1.0 \, \mathrm{s}$, which are finally averaged. Four such measurements are subsequently conducted at the four spectrometer positions. Due to drift of the sample and the diffraction pattern, contaminations \citep{Mitchell2015}, and beam damage \citep{Egerton2004}, the intensities of the four total spectra are slightly different. To account for these inequalities, the spectra are linearly aligned within energy windows below and above the absorption edges. For this purpose the sum spectrum at position `$-+$' is divided by each of the remaining three spectra. Then the three resulting quotients are fitted by linear functions thereby neglecting the energy windows of the Fe $L_3$ and $L_2$ absorption edges. Multiplying all but the `$-+$' spectra with these linear regressions results in an alignment of the intensities of all four sum spectra. The likewise mutually aligned EEL spectra are displayed in Figure \ref{EEL_spectra_exp} with the corresponding EMCD signals and positions of the spectrometer entrance aperture shown as insets. As expected from the theory of EMCD \citep{Hebert2003}, EEL spectra from neighbouring detector positions (top and bottom panels of Figure \ref{EEL_spectra_exp}) show a distinct asymmetry at the Fe absorption edges, which manifests itself in a difference of the intensities, the sign of which is opposite for the $L_3$ to the $L_2$ edges, respectively. This imbalance, however, disappears when spectra obtained from diagonally opposed aperture positions are compared (middle panel of Figure \ref{EEL_spectra_exp}).

\subsection{Quantifying $m_l/m_s$}

\begin{figure}[p]
        \centering
        \includegraphics[width=\textwidth]{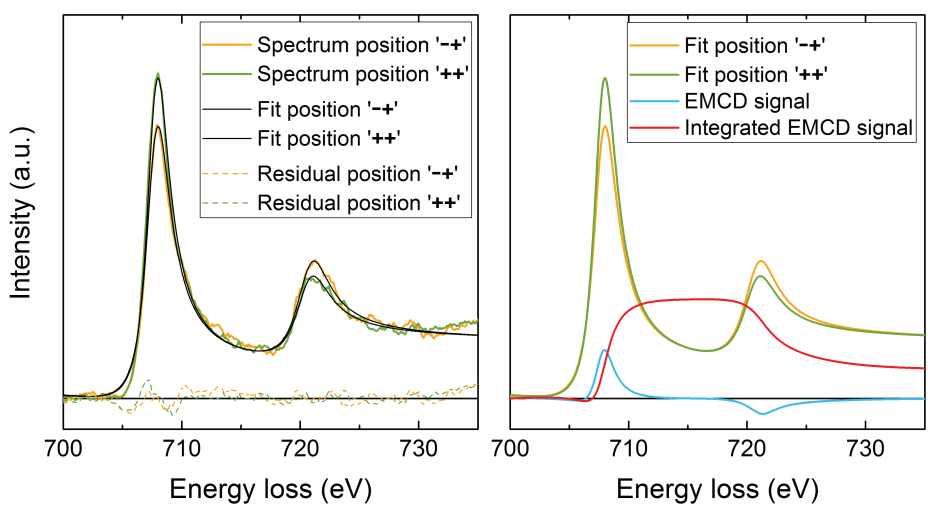}
        \caption{Left: Deconvolved EEL spectra at the detector positions `$-+$' (orange) and `$++$' (green). The fitted spectra are displayed as solid black lines. Furthermore the residua are plotted as dashed lines with the respective colour. Right: Same fitted spectra as on the left. The EMCD signal (blue) is determined as the difference between those two spectra. Integrating the EMCD signal over an energy window covering the absorption edges results in the integrated EMCD signal, which is used for the quantitative analysis. }
        \label{EEL_Fit}
\end{figure}

In order to retrieve quantitative information from the spectra, further processing of the experimental data is needed. This data treatment is explained as an example for the first two recorded spectra, namely `$-+$' and `$++$', since those contain the least artefacts caused by contamination and/or beam damage. In a first step, in order to remove artifacts due to multiple scattering, the spectra are deconvoluted with a low loss spectrum acquired under identical experimental conditions as the core loss spectra. For this deconvolution, the Richardson-Lucy algorithm is used \citep{Richardson1972,Lucy1974,Schmit2007}. After 15 iterations the deconvolution converges and yields a clearly visible smoothing of the spectra (see green and orange lines in the left part of Figure \ref{EEL_Fit}). Despite the extensive post-experimental analysis of the EEL spectra, the SNR is not yet enhanced to an extent that a smooth integrated EMCD curve can be extracted. Further improvement is thus achieved by fitting the de-convolved spectra with the fit function introduced in the previous part (see black lines in the left part and green and orange lines in the right part of Figure \ref{EEL_Fit}).

The likewise derived smooth fit functions (orange and green lines in the right part of Figure \ref{EEL_Fit}) are now used to calculate the EMCD signal  (blue line). Integration of this EMCD signal over the energy yields the red curve, which is used to quantify the magnetism of the nanoparticles. Applying the sum rules \citep{Rusz2007}, the ratio of the orbital to spin magnetic moments for the Fe atoms in the L$1_0$ ordered FePt nanoparticle is determined to be ${m_l}/{m_s} = 0.08 \pm 0.02$. The uncertainty of ${m_l}/{m_s}$ is calculated from the Poisson error of the total counts (de-convolved EEL spectra including the background), the error arising from the background removal, and the error due to the linear alignment of the spectra. The uncertainty due to the counting statistics is estimated from the square root of each intensity value of the de-convolved spectra (including the background). For the removal of the background, the pre-edge spectrum is fitted by a power law function $f\left(E\right) = AE^{-r}$ in the energy range from $687.7\, \mathrm{eV}$ to $704.6 \, \mathrm{eV}$. With the resulting fitting parameters $A$ and $r$ and their errors, the statistical uncertainty of the background fit is calculated for an energy range from $700\, \mathrm{eV}$ to $737 \, \mathrm{eV}$, which covers the Fe absorption edges relevant for the quantification process. The calculation is performed following the description of Egerton \citep{Egerton2011}. The uncertainty due to the linear alignment are taken from the statistical errors of the linear regressions in the alignment process. All uncertainties  are summed in quadrature and propagated through the sum rules. The limited precision of the final fitting procedure becomes apparent from the residuals of the fitting functions (cf. dashed lines in Figure \ref{EEL_Fit}). However, since these fitting functions model the scattering experimental data, they basically represent an averaging procedure. This becomes apparent from the fact that an integration of the EMCD signal as calculated from the difference between the un-smoothened `$-+$' and `$++$' spectra does not provide for a smooth integral curve with a clear maximum and lowered plateau (not shown here), and as a consequence the therefrom derivable values for ${m_l}/{m_s}$ vary between a minimum of ${m_l}/{m_s} = 0.066$ and a maximum of ${m_l}/{m_s} = 0.10$. These values, however, lie exactly in the uncertainty band of the above result of ${m_l}/{m_s} = 0.08 \pm 0.02$.

Although it is obvious from Figure \ref{EEL_spectra_exp} that in agreement with the theoretical expectations, a qualitative EMCD effect can be observed on all neighboring positions of the spectrometer entrance apertures, the procedure demonstrated here fails for spectra that are measured after a longer time of exposure of the sample to the electron beam (not shown here). Apparently then, detrimental effects due to contamination, beam damage, and sample or orientation drift are growing too large. The latter's influence on the quantification of ${m_l}/{m_s}$ needs further research in the future.

A comparison of the value for ${m_l}/{m_s} = 0.08 \pm 0.02$ as determined in the present work with the results of XMCD experiments on ensembles of L$1_0$ ordered FePt nanoparticles reveals a very good agreement. Antoniak et al. report values of ${m_l}/{m_s} = 0.02$ for plasma treated (hence at least partially disordered) nanoparticles and ${m_l}/{m_s} = 0.09$ for annealed nanoparticles \citep{Antoniak2006}, while Dupuis et al. have measured  ${m_l}/{m_s} = 0.11$ for unprocessed nanoparticles and ${m_l}/{m_s} = 0.14$ for annealed nanoparticles \citep{Dupuis2015}. The general tendency of slightly larger values for ${m_l}/{m_s}$ for L$1_0$ ordered FePt nanoparticles when obtained from XMCD measurements can be explained by the way of measuring the absorption signal during the XMCD experiments. There, the absorption signal was always determined by measuring the total electron yield, which does not provide for a direct measurement of the absorbed X-rays, but rather the photoelectrons generated through the X-ray absorption process are detected \citep{Gudat1972}. The latter, however, is very surface sensitive with a probing depth of only about 2 nm to 3 nm \citep{Abbate1992}. In this near-surface region, the quenching of the orbital angular momentum is suppressed because of the (surface-induced) breaking of the (crystal) symmetry. The hereby enhanced orbital moment results in larger values of ${m_l}/{m_s}$ in the vicinity of the particle surface \citep{Antoniak2011}. In contrast, the EMCD signal is exclusively derived from the energy losses of fully transmitted electrons. It is thus much more volume sensitive and averages over the orbital moments of the whole particle, which in turn is expected to result in comparably smaller values of ${m_l}/{m_s}$.

\section{Conclusions}

EMCD measurements are at the verge of becoming an established method to study the magnetic properties of materials with highest lateral resolution in a transmission electron microscope both quantitatively and with element-specificity. So far, however, quantitative studies were limited to more or less extended thin films, while studies on smaller, e.g., nanoparticulate materials were limited to rather qualitative investigations. In the present paper, based on a thorough analysis of the typically poor signal-to-noise ratio, a combination of both simulation and experimental methods is introduced that allows to enhance the SNR in the EMCD spectra to an extent that a quantitative characterization of magnetic properties becomes feasible even in individual nanoparticles.

It was shown that in order to pave the way towards quantitative EMCD measurements at smallest length scales, extensive preliminary studies are mandatory. Simulations on the (elastic and inelastic) interaction of the electron beam with the sample are used to identify optimal conditions for the actual experiments. Most importantly, the thinning of the sample to an optimal thickness and the correct positioning of the EEL spectrometer entrance aperture provide for maximal signal strengths and in turn enhance the SNR. During the experiments, binned gain averaging of the EEL spectra \citep{Bosman2008} and the acquisition of high quality dark references \citep{Hou2009} help to significantly reduce the correlated noise of the CCD camera. After removal of multiple scattering artefacts by de-convolution using the Richardson-Lucy algorithm  \citep{Richardson1972,Lucy1974,Schmit2007} and fitting of the spectra, it is finally possible to quantify the magnetism of a single L$1_0$ ordered FePt nanoparticle for the first time using EMCD. The determined ratio of orbital to spin magnetic moments ${m_l}/{m_s} = 0.08 \pm 0.02$ of the here investigated FePt particle agrees well with the results of XMCD measurements on large ensembles of such FePt nanoparticles. The quality of the measurement even allows for a discussion of minor discrepancies between the results of these two methods, which can be attributed to a higher surface sensitivity in case of XMCD when measured via electron yield.

However there are still issues to overcome prior to establishing EMCD as a more routine technique to study magnetism at the nanoscale. Especially artefacts due to a (partial) oxidation of metallic samples and the effect of contaminations on the EMCD spectra are yet to be better understood. Further improvement of the SNR is also expected to come about with automated acquisition procedures that could help to reduce the overall exposure time of the sample to the electron beam. Given this potential for improvement, EMCD may at last help to unveal magnetic details at unrivaled spatial resolution.

\section{Acknowledgements}

We thank Olav Hellwig and Sung Hun Wee for providing to us the FePt sample. The authors are indebted to Almut P\"{o}hl and Tina Sturm for the preparation of the TEM sample. Stefan L\"{o}ffler acknowledges financial support by the Austrian Science Fund (FWF) under grant nr. J3732-N27. J\'{a}n Rusz acknowledges Swedish Research Council, STINT and G\"{o}ran Gustafsson's Foundation for financial support. Peter Schattschneider acknowledges financial support by the Austrian Science Fund (FWF) under grant nr. I543-N20.





\biboptions{sort&compress}
\bibliographystyle{model1-num-names}







\clearpage

\clearpage

\newpage

\clearpage

\clearpage

\clearpage

\clearpage

\newpage

\clearpage

\end{document}